\documentclass[12pt]{article}
\usepackage{amsfonts}
\usepackage{amsmath}
\usepackage{cite}

\csname @addtoreset\endcsname{equation}{section}
\newcommand{\eq}{\begin{equation}}
\newcommand{\feq}{\end{equation}}
\newcommand{\eqn}{\begin{eqnarray}}
\newcommand{\feqn}{\end{eqnarray}}
\newcommand{\arr}{\begin{eqnarray*}}
\newcommand{\farr}{\end{eqnarray*}}

\newcommand{\bide}{\overleftarrow{\partial}\!\!\!\times\!\!\!\overrightarrow
{\partial}}

\font\mybb=msbm10 at 12pt
\def\bb#1{\hbox{\mybb#1}}

\def\bR {\bb{R}}

\begin{document}

\begin{titlepage}
\begin{flushright}
IFUM-708-FT \\
CAMS/02-03 \\
hep-th/0203038
\end{flushright}
\vspace{.3cm}
\begin{center}
\renewcommand{\thefootnote}{\fnsymbol{footnote}}
{\Large \bf Noncommutative Gravity in two Dimensions}
\vskip 15mm
{\large \bf {S.~Cacciatori$^{1,4}$\footnote{cacciatori@mi.infn.it},
A.~H.~Chamseddine$^2$\footnote{chams@aub.edu.lb},
D.~Klemm$^{3,4}$\footnote{dietmar.klemm@mi.infn.it},
L.~Martucci$^{3,4}$\footnote{luca.martucci@mi.infn.it},
W.~A.~Sabra$^2$\footnote{ws00@aub.edu.lb},
and D.~Zanon$^{3,4}$\footnote{daniela.zanon@mi.infn.it}}} \\
\renewcommand{\thefootnote}{\arabic{footnote}}
\setcounter{footnote}{0}
\vskip 10mm
{\small
$^1$ Dipartimento di Matematica dell'Universit\`a di Milano, \\
Via Saldini 50, I-20133 Milano. \\

\vspace*{0.5cm}

$^2$ Center for Advanced Mathematical Sciences (CAMS) and \\
Physics Department, American University of Beirut, Lebanon. \\

\vspace*{0.5cm}

$^3$ Dipartimento di Fisica dell'Universit\`a di Milano, \\
Via Celoria 16, I-20133 Milano. \\

\vspace*{0.5cm}

$^4$ INFN, Sezione di Milano, \\
Via Celoria 16,
I-20133 Milano. \\
}
\end{center}
\vspace{2cm}
\begin{center}
{\bf Abstract}
\end{center}
{\small We deform two-dimensional topological gravity
by making use of its gauge theory formulation. The obtained
noncommutative gravity model is shown to be invariant under
a class of transformations that reduce to standard diffeomorphisms
once the noncommutativity parameter is set to zero. Some
solutions of the deformed model, like fuzzy AdS$_2$, are obtained.
Furthermore, the transformation properties of the model under
the Seiberg-Witten map are studied.}

\end{titlepage}

\section{Introduction}

Quantum field theories formulated on a noncommutative space have
become recently the object of renewed interest. This is primarily due to
the realization that noncommutative $U(N)$ gauge theories arise in the
field theory limit of strings in a constant $B$-field
background \cite{Seiberg:1999vs}.
The perturbative analysis of such theories is simplified by the fact
that the noncommutativity of spacetime can be traded off by a 
modified multiplication rule for the fields, i.e. functions on
noncommutative $\mathbb{R}^d$ can be treated as ordinary functions on
standard $\mathbb{R}^d$ with a deformed multiplication
given by the Moyal $*$-product.

By now there exists an abundant literature on the perturbative and non
perturbative studies of field theories in noncommutative {\em flat}
$\mathbb{R}^d$. For a review and an extensive list of references
cf.~e.~g.~\cite{Douglas:2001ba}.
On the contrary, very little is known about corresponding
theories in a {\em curved} noncommutative
space or else about a noncommutative formulation of gravity
itself. One of the main obstacles to overcome in the formulation of
gravity on noncommutative spaces is related to the fact that the Moyal
product does not maintain reality. One possible way to preserve
nevertheless reality of the gravitational fields is to use explicitely
the Seiberg-Witten map \cite{Chamseddine:2000si}. Otherwise it seems that one
is forced to complexify the
fields \cite{Chamseddine:2000zu,Chamseddine:2000rj,Nishino:2001gt}. However,
complex gravity may be plagued by inconsistencies already at the
commutative level \cite{Damour:1992bt,Moffat:fc}.

Here we attack the problem starting from the theory of noncommutative gravity
formulated in two space-time dimensions. In this case we can take advantage
of our knowledge about non commutative gauge theories. There we know how
to deform the gauge transformations for the fields and everything is under
control at least at the kinematical level. In fact the Jackiw-Teitelboim
(JT) model \cite{Teitelboim:ux} of $2$-$d$ commutative
gravity can be written as a topological $SU(1,1)$ gauge
theory \cite{Isler:1989hq,Chamseddine:1990wn}.
Within this formulation embedding in a noncommutative space is
straightforward: since in the volume form the metric does not appear it
is sufficient to introduce the $*$-product appropriately and
extend the gauge group to $U(1,1)$\footnote{For other attempts to formulate
gravity on noncommutative spaces based on Chern-Simons actions and its
variants cf.~\cite{Banados:2001xw,Nair:2001kr,Cacciatori:2002gq,
Shiraishi:2002qu,Chamseddine:2002fd}.}.
We will show how to write the action
in terms of real fields and how to achieve the decoupling of the extra
$U(1)$ in the commutative limit. Then we address the issue of diffeomorphisms.
We find that the deformed action is invariant under a class of transformations
that reproduce the standard diffeomorphisms in the commutative limit.
Moreover, once the equations of motion are imposed, they
are equivalent to gauge transformations, as it happens in the commutative
case.

Our paper is organized as follows: in the next section we define the
noncommutative gravity action in terms of a topological two-dimensional
gauge theory and write the equations of motion. In section 3
it is shown that the Seiberg-Witten (SW) formula
maps the deformed model into the standard commutative topological gauge theory.
In section 4 we show
that the action enjoys an invariance that reduces
to ordinary diffeomorphism invariance in the commutative case.
In section 5
we obtain solutions of the equations of motion and discuss their dependence
on the noncommutativity parameter. Finally we
present our conclusions.

\section{Deformation of two-dimensional topological gauge theory}

It is well-known that the JT model \cite{Teitelboim:ux} of dilaton gravity in
two dimensions can be formulated as an $SU(1,1)$ topological gauge
theory \cite{Isler:1989hq,Chamseddine:1990wn}.
This is similar to the three-dimensional case,
where pure Einstein gravity can be written as a Chern-Simons
theory \cite{Achucarro:1986vz}. In what follows, we will use this gauge theory
formulation to define $2$-$d$ noncommutative gravity. To this end, we first
note that the group $SU(1,1)$ is not closed with respect to the Moyal product,
and thus we are forced to consider a gauge theory based on $U(1,1)$.
The action of the $U(1,1)$ gauge theory on noncommutative $\bR^2$,
with coordinates $x^{\mu}$ satisfying $[x^{\mu}, x^{\nu}] = i\theta^{\mu\nu}$,
reads\footnote{Of course in two dimensions
$[x^{\mu}, x^{\nu}] = i\theta^{\mu\nu}$ implies timelike noncommutativity,
so one might think that problems like causality violation or loss of
unitarity \cite{Seiberg:2000gc,Gomis:2000zz} occur. However, as we will see
below (cf.~Eq.~(\ref{F=0})), the deformed theory is still topological,
so that there are no propagating degrees of freedom, and we thus expect
the theory to be well-defined in spite of timelike noncommutativity.}

\begin{equation}
S = \beta \int {\mathrm{Tr}} (\Phi \star F)\,, \label{actiongauge}
\end{equation}

where $\beta$ is a dimensionless coupling constant. 
$\Phi = \Phi^A \tau_A$, $A=A^A_\mu\tau_Adx^\mu$ and 
$F = \frac 12 F^A_{\mu\nu}\tau_A dx^{\mu} \wedge
dx^{\nu}$ take values in the Lie algebra $u(1,1)$ (the generators $\tau_A$
are given in the appendix). The star denotes the usual Moyal product,
and

\begin{eqnarray}
F &=& DA=dA+A\wedge^\star A\ ,\cr
&\Downarrow& \cr
F_{\mu\nu} & =& \partial_{\mu} A_{\nu} - \partial_{\nu} A_{\mu}
+ A_{\mu} \star A_{\nu} - A_{\nu} \star A_{\mu}\,.
\end{eqnarray}

Note that the fields $\Phi^A$, $A^A_{\mu}$ and $F^A_{\mu\nu}$
are real, and that this reality is preserved under the infinitesimal gauge
transformations\footnote{In what follows, all commutators and
anticommutators are taken with respect to the Moyal product.}

\begin{equation}
\delta_\lambda  A = d\lambda +[A,\lambda]=D\lambda\,, \qquad
\delta_\lambda \Phi = [\Phi , \lambda]\,,
\label{gaugetransformations}
\end{equation}

where $\lambda=\lambda^A\tau_A$. The integrated form of
(\ref{gaugetransformations}) is given by

\begin{equation}
A \to g^{-1}_\star \star A \star g + g^{-1}_\star \star dg\,, \qquad 
\Phi \to g^{-1}_\star \star \Phi \star g\,,
\end{equation}

with $g=\exp_\star \lambda \in U(1,1)_{\star}$, i.~e.~,
$g_{\star}^{-1} = \eta g^{\dag} \eta$, where $\eta = {\mathrm{diag}}(-1,1)$
and $g \star g_{\star}^{-1} = 1 = g_{\star}^{-1} \star g$.\\
The equations of motion following from (\ref{actiongauge}) read

\begin{eqnarray}
F &=& 0\,, \label{F=0} \\
D \Phi &=& d \Phi + [A, \Phi] = 0\,,
                       \label{DPhi=0}
\end{eqnarray}

so that the solutions are those of flat $U(1,1)$ connections in
two dimensions. The equation (\ref{F=0}) implies that locally one can write

\begin{equation}
A_{\mu} = g^{-1}_\star \star \partial_{\mu} g\,, \label{puregauge}
\end{equation}

with $g \in U(1,1)_{\star}$. Using

\begin{equation}
\partial_{\mu} g = i\theta_{\mu\nu} [g, x^{\nu}]\,,
\end{equation}

where $\theta_{\mu\nu}\theta^{\nu\lambda} = \delta_{\mu}^{\lambda}$,
and defining the covariant coordinates

\begin{equation}
X^{\mu} = x^{\mu} + i\theta^{\mu\nu} A_{\nu}\,,
          \label{covcoord}
\end{equation}

one immediately obtains

\begin{equation}
X^{\mu} = g^{-1}_\star \star x^{\mu} \star g\,.
           \label{resultX}
\end{equation}

Note that the $X^{\mu}$ satisfy $[X^{\mu}, X^{\nu}] = i\theta^{\mu\nu}$.
We still have to solve Eq.~(\ref{DPhi=0}), which is equivalent to
$[\Phi, X^{\mu}] = 0$. Inserting (\ref{resultX}) and the ansatz
$\Phi = g^{-1} \star B \star g$, where $B \in u(1,1)$, one gets
$\partial_{\mu} B = 0$, so $B$ is constant. The solution of
(\ref{DPhi=0}) is thus

\begin{equation}
\Phi = g^{-1}_\star \star B \star g\,. \label{solPhi}
\end{equation}

In order to make contact with gravity,
we decompose the $u(1,1)$ valued scalar and gauge fields according to

\begin{equation}
\Phi^A = (\phi^a, \phi, \rho)\,, \qquad A^A_{\mu} = (e^a_{\mu}/l, \omega_{\mu},
         b_{\mu})\,, \label{decomp}
\end{equation}

where $a = 0,1$ and $l$ is related to the negative cosmological constant
by $\Lambda = -1/l^2$. We note that in the noncommutative
case one is forced to
include the fields $\rho$ and $b_{\mu}$, which correspond to the trace
part of u(1,1), in addition to the usual spin
connection $\omega_{\mu} = \omega_{01\mu}$, the zweibein $e^a_{\mu}$,
and the scalars $\phi$ and $\phi^a$.\\
In what follows it will be convenient to define

\begin{eqnarray}
\Omega^a_{\mu\, b} &=& \epsilon^a_{\,\,\,b} \omega_{\mu} + i \delta^a_{\,\,\,b}
                       b_{\mu}\,, \label{NCspinconn} \\
T^a_{\mu\nu} &=& \partial_{\mu} e^a_{\nu} - \partial_{\nu} e^a_{\mu}
                 + \frac 12 [\Omega^a_{\mu\,b}, e^b_{\nu}]
                 - \frac 12 [\Omega^a_{\nu\,b}, e^b_{\mu}]\,, \label{NCtors} \\
\phi_{ab} &=& \phi \epsilon_{ab} - i\rho \eta_{ab}\,, \label{NCscal}
\end{eqnarray}

where

\begin{equation}
[\Omega^a_{\mu\,b}, e^b_{\nu}] \equiv \Omega^a_{\mu\,b} \star e^b_{\nu} -
e^b_{\nu} \star \Omega_{\mu\,b}^{\quad a}\,.
\end{equation}

We will see below that $\Omega^a_{\,\,\,b}$ can be interpreted as a
noncommutative $so(1,1) \oplus u(1)$ spin connection, with a
trace part given by the abelian gauge field $b_{\mu}$, whereas $T^a$
is the noncommutative torsion.\\
Using the decomposition (\ref{decomp}) and (\ref{NCspinconn})
- (\ref{NCscal}), the action (\ref{actiongauge}) can be written as

\begin{equation}
S = \frac{\beta}{4} \int d^2x \,\epsilon^{\mu\nu} \left[\phi_{ab} \star
    \left({\cal R}^{ab}_{\mu\nu} - 2 \Lambda e^a_{\left[\mu\right.}
    \star e^b_{\left. \nu\right]}\right) - 2\phi_a \star T^a_{\mu\nu}\right]\,,
    \label{actiongrav}
\end{equation}

where we defined the noncommutative curvature two-form

\begin{equation}
{\cal R}^{ab}_{\mu\nu} = \partial_{\mu}\Omega_{\nu}^{ab} -
                         \partial_{\nu}\Omega_{\mu}^{ab} + \frac 12
                         [\Omega^a_{\mu\, c}, \Omega^{cb}_{\nu}] - \frac 12
                         [\Omega^a_{\nu\, c}, \Omega^{cb}_{\mu}]\,,
\end{equation}

from which we can see that indeed $\Omega^a_{\,\,\,b}$ plays the role
of an $so(1,1) \oplus u(1)$ spin connection. The action (\ref{actiongrav})
defines our model of noncommutative gravity in two dimensions.\\
As in the commutative case, the fields $\phi^a$ are Lagrange multipliers
imposing the constraint $T^a_{\mu\nu} = 0$, i.~e.~, the vanishing of
the noncommutative torsion.\\
The other equations of motion following from
(\ref{actiongrav}) read

\begin{eqnarray}
{\cal R}^a_{\;\; b\,\mu\nu} - \frac{1}{2l^2} \epsilon^a_{\,\,\,b}
\epsilon_{cd}\{e^c_{\mu}, e^d_{\nu}\} + \frac{1}{2l^2}
\delta^a_{\,\,\,b}\eta_{cd}
[e^c_{\mu}, e^d_{\nu}] &=& 0\,, \nonumber \\
\partial_{\nu}\phi_{ab} + \frac 12 \left(\Omega^c_{\nu\, a} \star \phi_{cb} -
\phi_{ac} \star \Omega_{\nu\, b}^{\quad c}\right)
- \frac 12 \epsilon_{ab}\epsilon_{cd}\{e^c_{\nu},
\phi^d\} - \frac 12 \eta_{ab}\eta_{cd}[e^c_{\nu}, \phi^d] &=& 0\,, \\
\partial_{\nu}\phi^a + \frac 12 \left(\Omega^a_{\nu\,b} \star \phi^b -
\phi^b \star \Omega_{\nu\, b}^{\quad a}\right) + \frac{1}{2l}\left(
\phi_b^{\,\,\,a} \star e^b_{\nu} - e^b_{\nu} \star \phi^a_{\,\,\,b}\right)
&=& 0\,. \nonumber
\end{eqnarray}

Now one can construct a metric according to

\begin{equation}\label{metric}
G_{\mu\nu} = e^a_{\mu} \star e^b_{\nu}\, \eta_{ab} =
             g_{\mu\nu} + iB_{\mu\nu}\,,
\end{equation}

where

\begin{equation}\label{smetric}
g_{\mu\nu} = \frac 12 \eta_{ab} \{e^a_{\mu}, e^b_{\nu}\}
\end{equation}

is real and symmetric, and reduces to the usual expression for the
metric in the commutative case, whereas

\begin{equation}\label{ametric}
B_{\mu\nu} = -\frac i2 \eta_{ab} [e^a_{\mu}, e^b_{\nu}]
\end{equation}

is real and antisymmetric, and vanishes for $\theta^{\mu\nu} = 0$.

We finally note that the action (\ref{actiongauge}) can be rewritten
as a matrix model for the covariant coordinates $X^{\mu}$ defined
in (\ref{covcoord}).
Setting $\theta = \theta^{01}$, the action reads

\begin{equation}
S = -\frac{\beta}{2\theta^2} \epsilon_{\mu\nu}{\mathrm{Tr}}_{{\cal H}}
    (\Phi \star ([X^{\mu}, X^{\nu}] - i\theta^{\mu\nu}))\,.
    \label{matrixmodel}
\end{equation}

Here $\Phi$ and $X^{\mu}$ are operators acting on a Hilbert space
${\cal H} = {\cal H}' \otimes V$, where ${\cal H}'$ is the
infinite-dimensional subspace of the Hilbert space carrying the
irreducible representation of $[x^{\mu}, x^{\nu}] = i\theta^{\mu\nu}$,
whereas $V$ carries the fundamental representation of
$u(1,1)$ \cite{Douglas:2001ba}.

From (\ref{matrixmodel}) we see that
the scalar $\Phi$ is a Lagrange multiplier enforcing noncommutativity
of the covariant coordinates,

\begin{equation}
[X^{\mu}, X^{\nu}] = i\theta^{\mu\nu}\,,
\end{equation}

and $\theta$ enters the coupling constant. It is interesting to note
that, by generalizing the map from
$u(2)$ to $u(1)$ gauge models constructed in
\cite{Nair:2001rt}\footnote{Cf.~also~\cite{Kiritsis:2002py}.}
to a correspondence between $u(1,1)$ and $u(1)$, one can represent the
$u(1,1)$ valued functions $\Phi$ and $X^{\mu}$ as scalar functions
in a $u(1)$ theory\footnote{Note however that due to the modified group
metric in our case ($U(1,1)$ rather than $U(2)$) this mapping is less
trivial than the one in \cite{Nair:2001rt}, and would produce a nonstandard
$U(1)$ action. We would like to thank A.~Polychronakos for comments on this.}.

\section{The Seiberg-Witten map}

In \cite{Grandi:2000av} it was shown that the Seiberg-Witten formula
maps the Chern-Simons (CS) action on noncommutative spaces into the
standard commutative Chern-Simons action. As the commutative version
of the topological gauge theory (\ref{actiongauge}) can be obtained from
the CS theory by dimensional reduction, one might ask whether
a similar property holds for (\ref{actiongauge}), i.~e.~, whether it is
related to the standard commutative topological gauge theory by
the Seiberg-Witten map. We will now show that this is indeed
the case.

A correspondence between commutative and noncommutative gauge field
theories can be defined by the SW
map \cite{Seiberg:1999vs}\footnote{For an alternative derivation
of the SW equation cf.~\cite{Bichl:2001yf}.}

\begin{eqnarray}
\delta \Phi &=& -\frac i4 \delta\theta^{\alpha\beta}\{A_{\alpha},
                \partial_{\beta}\Phi + D_{\beta}\Phi\}\,, \nonumber \\
\delta F_{\mu\nu} &=& \frac i4 \delta\theta^{\alpha\beta}\left[
                      2\{F_{\mu\alpha}, F_{\nu\beta}\} - \{A_{\alpha},
                      \partial_{\beta}F_{\mu\nu} + D_{\beta}F_{\mu\nu}\}
                      \right]\,, \label{SWmap}
\end{eqnarray}

where the transformation formula for the adjoint scalar $\Phi$ can be
found by dimensional reduction from three to two dimensions, setting
$\theta^{2\mu} = 0, \mu = 0,1$, and $\Phi = A_3$.

In order to study the variation of the action (\ref{actiongauge})
under the SW map, we differentiate it with respect to $\theta^{\alpha\beta}$,

\begin{equation}
\frac{\delta S}{\delta \theta^{\alpha\beta}} = -\frac{\beta}{2}\int d^2x
\epsilon^{\mu\nu}{\mathrm{Tr}}\left(\frac{\delta F_{\mu\nu}}
{\delta \theta^{\alpha\beta}} \star \Phi + F_{\mu\nu} \star \frac{\delta \Phi}
{\delta \theta^{\alpha\beta}}\right)\,.
\end{equation}

Using (\ref{SWmap}), this simplifies after some algebra to

\begin{equation}
\frac{\delta S}{\delta \theta^{\alpha\beta}} = \frac{\beta}{2}\int d^2x
{\mathrm{Tr}}\left[-\frac i4 \{F_{\alpha\beta}, \Phi\} \star F_{01}
-\frac i4 \Phi \star \{F_{0\alpha}, F_{\beta 1}\} + \frac i4 \Phi \star
\{F_{0\beta}, F_{\alpha 1}\}\right]\,.
\end{equation}

As we are in two dimensions, the only nonvanishing component of
$\theta^{\mu\nu}$ is $\theta^{01} = \theta$. It is then easy to see that

\begin{equation}
\frac{\delta S}{\delta \theta} = 0\,, \label{deltaS}
\end{equation}

and therefore the deformed action is mapped to the standard commutative
one. We have to keep in mind, however, that the SW map is of perturbative
nature in $\theta$, so the equivalence between the deformed and undeformed
models holds perturbatively. Of course, the noncommutative theory
(\ref{actiongauge}) can also have solitonic solutions that become singular
for $\theta \to 0$, and thus have no analogue in the commutative case.
Furthermore, in deriving (\ref{deltaS}), we discarded boundary terms,
so that in presence of a boundary the equivalence of the noncommutative
and the commutative topological gauge theory fails. This is analogous to the
case of Chern-Simons theory in three dimensions \cite{Grandi:2000av}.

\section{Gauge symmetries and deformed diffeomorphisms}

In this section we want to find a candidate for diffeomorphisms in the
noncommutative case\footnote{For related work cf.~\cite{Jackiw:2001jb}.}.
Since in the commutative limit we want to obtain
standard results, we start from what is known in that case. The commutative
version of our action
(\ref{actiongauge}) is invariant not only under the commutative version of the
gauge transformations (\ref{gaugetransformations}),
but also under infinitesimal diffeomorphisms (Lie derivatives)
along an arbitrary vector field $v=v^\mu \partial_\mu$,

\begin{eqnarray}\label{diff}
{\cal L}_v A &=& (di_v + i_v d)A\ ,\cr
             & & \cr {\cal L}_v \Phi &=&i_vd\Phi\ ,
\end{eqnarray}

\noindent where $i_v$ is the inner product on differential forms.

Using the Leibnitz rule for the inner product ($\omega^p$ and $\xi^q$ are 
respectively a p- and a  q-form),

\begin{equation} \label{Leibnitz}
i_v(\omega^p\wedge \xi^q) = (i_v \omega^p)\wedge \xi^q +(-)^p
                            \omega^p\wedge(i_v \xi^q)\ ,
\end{equation}

it is easy to prove that

\begin{eqnarray} \label{relations}
{\cal L}_v A &=& \delta_{i_v A}A  +i_v F\,, \cr
             & & \cr {\cal L}_v \Phi  &=& \delta_{i_v A}\Phi +i_vD\Phi\,.
\end{eqnarray}

From Eq.~(\ref{relations}) and from the equations of motion 
(\ref{F=0}-\ref{DPhi=0})
we see that, on-shell, the infinitesimal diffeomorphisms can be
written as gauge transformations with parameters $\lambda=i_v A$. This fact
can be used to
relate translations and gauge transformations with $\lambda=\alpha^a\tau_a$.
To this end we divide the connection into
a part containing the zweibein and a part containing the spin connection
and the center $U(1)$,
\begin{equation}
A=l^{-1}e+\Theta\ , \ e=e^a_\mu\tau_a dx^\mu\ ,\ \Theta=\omega_\mu \tau_2
dx^\mu + b_\mu \tau_3 dx^\mu \,.
\end{equation}

\noindent In this way, with an invertible
$e^a_\mu$, we can write

\begin{equation}
v^\mu:= le^\mu_a \alpha^a \Leftrightarrow \alpha^a=l^{-1}i_v e^a \,,
\end{equation}

\noindent thus obtaining\footnote{We
use the symbol $\doteq$ for equations which are valid on-shell.}

\begin{eqnarray}\label{trasf}
\delta_{\alpha^a \tau_a}A &\doteq& {\cal L}_v A + \delta_{i_v \Theta}A \ ,\cr
\delta_{\alpha^a \tau_a}\Phi &\doteq& {\cal L}_v \Phi +
\delta_{i_v \Theta}\Phi \ .
\end{eqnarray}

In the noncommutative case the situation is quite different. Whereas
the action (\ref{actiongauge}) is invariant under gauge transformations 
(\ref{gaugetransformations}),
the invariance under diffeomorphisms seems to be completely destroyed. 
However we will show that the results obtained in the commutative
case naturally suggest how to deform the diffeomorphism invariances in the
noncommutative case.

Let us introduce the most natural generalization of the inner product and the 
Lie derivative in the noncommutative theory. If $\omega^p$ is a p-form we
define

\begin{eqnarray}
\omega^p &=& \frac{1}{p!} \omega_{\mu_1 \ldots \mu_p} dx^{\mu_1}\wedge \ldots
\wedge dx^{\mu_p}\ ,\cr
&&\cr
i^\star_v\omega^p &:=& \frac{1}{2(p-1)!}[v^\rho \star 
\omega^p_{\rho \mu_1 \ldots \mu_{p-1}}+\omega^p_{\rho \mu_1 \ldots \mu_{p-1}}
\star v^\rho]\, dx^{\mu_1}\wedge \ldots \wedge dx^{\mu_{p-1}}\ ,\cr
&&\cr
{\cal L}^\star_v \omega^p &:=& \frac{1}{2p!}\left\{[v^\rho \star
\partial_\rho \omega^p_{\mu_1\ldots\mu_p}+(\partial_{\mu_1}v^\rho) \star
\omega^p_{\rho\mu_2\ldots\mu_p}+\ldots + (\partial_{\mu_p}v^\rho) \star
\omega^p_{\mu_1\ldots\mu_{p-1}\rho}]\right. \cr
&& + \left[(\partial_\rho \omega^p_{\mu_1\ldots\mu_p})\star  v^\rho +
\omega^p_{\rho\mu_2\ldots\mu_p}\star \partial_{\mu_1}v^\rho + \ldots \right.
\cr
&& + \left.\left.\omega^p_{\mu_1\ldots\mu_{p-1}\rho}\star \partial_{\mu_p}
v^\rho\right]\right\} dx^{\mu_1}\wedge \ldots
\wedge dx^{\mu_p} \cr
&& \cr
&=& (di^\star_v +i^\star_v d)\omega^p\ ,
\end{eqnarray}

\noindent where the exterior derivative $d$ is defined as in the commutative 
case and satisfies the same properties. It is important to note that the 
Leibnitz rule in Eq.~(\ref{Leibnitz}) is not valid anymore. Using the above
definitions we get

\begin{eqnarray} \label{ncLiebnitz}
i_v^\star[A\wedge^\star A] &=& (i^\star_v A)\star A-A\star
(i^\star_v A) + \frac 12
\big\{ A_\mu ,\big[ A,v^\mu\big] \big\}\ ,\cr
&&\cr
i_v^\star[\Phi\star A]&=& \Phi\star (i^\star_v A) - \frac 12
\big[ \Phi,v^\mu\big]\star A_\mu \ ,\cr
&&\cr
i_v^\star[A\star\Phi]&=&  (i^\star_v A)\star\Phi + \frac 12
A_\mu\star\big[ \Phi,v^\mu\big]\ .
\end{eqnarray}

Inspired by the commutative counterpart in (\ref{relations}) we define the
{\em deformed diffeomorphisms} as

\begin{eqnarray}\label{defdiff}
\Delta^\star_v&:=&i^\star_vD+\delta_{i^\star_v A}\ ,\cr
&\Downarrow&\cr
\Delta^\star_vA&=&i^\star_vF+\delta_{i^\star_v A}A\ ,\cr
\Delta^\star_v\Phi&=&i^\star_vD\Phi+\delta_{i^\star_v A}\Phi\ .
\end{eqnarray}

Now we want to prove that these transformations are symmetries of the action
modulo boundary terms. First we note that, if
$\xi$ is a 1-form and $\omega$ is a two form, a noncommutative
integrated version of the Leibnitz rule (\ref{Leibnitz}) is valid:

\begin{eqnarray} \label{intLeibnitz}
\int \xi \wedge^\star  i^\star_v \omega &=&  \int d^2x 
\epsilon^{\mu\nu} \xi_\mu \star (i^\star_v\omega)_\nu \cr
&=& \frac 12 \int d^2x \epsilon^{\mu\nu}\xi_\nu\star 
(v^\rho \star \omega_{\rho\nu} +
\omega_{\rho\nu}\star v^\rho) \cr
&=& \frac 12 \int d^2x \epsilon^{\mu\nu} (v^\rho \star \xi_\mu +
\omega_\mu \star v^\rho)\star \omega_{\rho\nu} \cr
&=& \frac 14 \int d^2x \epsilon^{\mu\nu} (v^\rho \star \xi_\rho +
\omega_\rho \star v^\rho)\star \omega_{\mu\nu} \cr
&=& \int (i^\star_v \xi)\star \omega\ .
\end{eqnarray}

\noindent Note that in order to obtain (\ref{intLeibnitz}) the symmetric
form of the definition of $i^\star$ and the fact that we are in
two dimensions are crucial.

Since we know that the action is gauge invariant 
under (\ref{gaugetransformations}), in order
to prove its invariance under the deformed diffeomorphisms (\ref{defdiff})
it is sufficient to show that it remains
unaltered by the transformations

\begin{eqnarray}
\delta^\prime_v A &=& i^\star_v F\ ,\cr
&&\cr
\delta^\prime_v \Phi &=& i^\star_v D\Phi\ .
\end{eqnarray}

\noindent Using Eq.(\ref{intLeibnitz}), we obtain

\begin{eqnarray}
\delta^\prime_v S&=& \beta\int Tr \big[ i^\star(D\Phi)\star F+
\Phi\star ( di^\star_v F + i^\star_v F\wedge^\star A +A\wedge^\star 
i^\star_v F)\big] \cr
&=& \beta \int Tr \big( D\Phi\wedge^\star i^\star_v F -d\Phi\wedge^\star
i^\star_v F +\Phi\star i^\star_v F\wedge^\star A +\Phi\star A\wedge^\star 
i^\star_v F \big) \cr
&=& \beta \int  Tr \big[ D\Phi\wedge^\star i^\star_v F -(d\Phi +A\star\Phi
-\Phi\star A)\wedge^\star 
i^\star_v F \big] \cr
&=& 0\,.
\end{eqnarray}

Thus we have shown that the deformed diffeomorphisms (\ref{ncdiff}) are
indeed symmetries of the action.

We can also write the noncommutative version of Eq.~(\ref{trasf}) as

\begin{eqnarray}
\delta_{\alpha^a \tau_a}A &\doteq& \Delta_v^\star A +
\delta_{i^\star_v \Theta}A \,, \cr
\delta_{\alpha^a \tau_a}\Phi &\doteq& \Delta^\star_v \Phi +
\delta_{i^\star_v \Theta}\Phi \,,
\end{eqnarray}     

\noindent where now $v=v^\mu\partial_\mu$ is such that
$\alpha^a=l^{-1}i^\star_v e^a$. Note that the vector $v$ can always 
be chosen to be real. This is shown in appendix \ref{vreal}.\\
Using Eq.(\ref{ncLiebnitz}), it is straightforward to prove that the deformed
diffeomorphisms (\ref{defdiff}) can be written as 

\begin{equation} \label{ncdiff}
\Delta^\star_v= {\cal L}^\star_v + \frac 12 \big\{ A_\mu,\big[\  .\;  ,v^\mu
\big]\big\}\,.
\end{equation}

The transformation properties of the fields (\ref{decomp})
under the action of (\ref{ncdiff}) are given in appendix \ref{transfprop}.
In particular, the symmetric part $g_{\mu\nu}$ and the antisymmetric
part $B_{\mu\nu}$ of the metric (\ref{metric}) transform as

\begin{eqnarray*}
\Delta_{v}^\star g_{\mu\nu} &=& {\cal L}_v^\star g_{\mu\nu} +
\frac 14\eta_{ab}([[e^a_\mu,v^\rho],\partial_\rho e^b_\nu] +
[[e^a_\nu,v^\rho],\partial_\rho e^b_\mu] \cr
&& + [[e^a_\mu,\partial_\nu v^\rho], e^b_\rho] +
[[e^a_\nu,\partial_\mu v^\rho], e^b_\rho]) \cr
&& + \frac 18 \epsilon_{ab}(\{e^a_\mu,[\omega_\rho,[e^b_\nu,v^\rho]]\}+
\{e^a_\nu,[\omega_\rho,[e^b_\mu,v^\rho]]\} \cr
&& - \{e^a_\mu,[e^b_\rho,[b_\nu,v^\rho]]\}-
\{e^a_\nu,[e^b_\rho,[b_\mu,v^\rho]]\})\,,\cr
&&\cr
\Delta_{v}^\star B_{\mu\nu} &=& {\cal L}_v^\star B_{\mu\nu}
- \frac i4 \eta_{ab}(\{[e^a_\mu,v^\rho],\partial_\rho e^b_\nu\}-
\{[e^a_\nu,v^\rho],\partial_\rho e^b_\mu\} \cr
&& + \{[e^a_\mu,\partial_\nu v^\rho], e^b_\rho\}-
\{[e^a_\nu,\partial_\mu v^\rho], e^b_\rho\}) \cr
&& - \frac{i}{8}\epsilon_{ab}([e^a_\mu,[\omega_\rho,[e^b_\nu,v^\rho]]]-
[e^a_\nu,[\omega_\rho,[e^b_\mu,v^\rho]]] \cr
&& - [e^a_\mu,[e^b_\rho,[b_\nu,v^\rho]]]+
[e^a_\nu,[e^b_\rho,[b_\mu,v^\rho]]])\,.
\end{eqnarray*}

One may note that these transformations reduce to ordinary diffeomorphisms
if $\theta^{\mu\nu} = 0$.

\section{Some solutions}

\subsection{Fuzzy AdS}

The undeformed gravitational action admits the AdS$_2$ solution

\begin{equation}
ds^2 = -\frac{r^2}{l^2}dt^2 + \frac{l^2}{r^2}dr^2\,. \label{metricr}
\end{equation}

This leads to the connection

\begin{equation}
A_t = \frac{ir}{2l^2} \left(\begin{array}{cc} 1 & 1 \\
      -1 & -1 \end{array} \right)\,, \qquad
A_r = \frac{1}{2r} \left(\begin{array}{cc} 0 & 1 \\ 1 & 0
      \end{array} \right)\,,
\end{equation}
which can be written as $A=g^{-1}dg$ with

\begin{equation}
g = \frac 12 \sqrt{\frac rl}\left(\begin{array}{cc}
                          1+\frac{it}{2l} & \frac{it}{2l} \\
                          -\frac{it}{2l} & 1-\frac{it}{2l} \end{array} \right) 
                          \left(\begin{array}{cc}
                          1+\frac{l}{r} & 1-\frac{l}{r}\\
                          1-\frac{l}{r} & 1+\frac{l}{r} \end{array} \right)\,.
\end{equation}

At this point, we observe that $g$ is the product of two matrices
$f,h\in SU(1,1)$, each one depending on a single variable only. This implies
that $f,h$ are elements of $U(1,1)_{\star}$.
Consequently we have  $\tilde g :=f\star h \in U(1,1)_{\star}$ and 
we can use $\tilde g$ to obtain a noncommutative solution 
$\tilde A_\mu =\tilde{g}_{\star}^{-1} \partial_\mu \tilde g$.\\
With these matrices, we easily obtain
\begin{equation}
\tilde A_\mu =A_\mu \,,
\end{equation}
so that (\ref{metricr}) is a solution of the noncommutative gravity. This is
the fuzzy AdS$_2$.\\
Fuzzy AdS$_2$ was obtained in \cite{Ho:2001br}\footnote{Cf.~also
\cite{Ho:2000fy}.} by an analytic continuation
of the fuzzy sphere \cite{Madore:1992bw}. Let us briefly recall the
construction. We denote the Cartesian coordinates of AdS$_2$ by
$X^{-1}, X^0, X^1$. The algebra of fuzzy AdS$_2$ is \cite{Ho:2001br}

\begin{eqnarray}
[X^{-1}, X^0] &=& -i\theta l^{-1} X^1\,, \nonumber \\
{[}X^0, X^1] &=& i\theta l^{-1} X^{-1}\,, \label{algebra} \\
{[}X^1, X^{-1}] &=& i\theta l^{-1} X^0\,, \nonumber
\end{eqnarray}

where $\theta$ is the noncommutativity parameter, and $l$
the curvature radius of AdS$_2$, so

\begin{equation}
\eta_{ij}X^i X^j = -(X^{-1})^2 - (X^0)^2 + (X^1)^2 = -l^2\,, 
\end{equation}

with $(\eta_{ij}) = {\mathrm{diag}}(-1,-1,1)$. The isometry group
$SU(1,1)$ of AdS$_2$ preserves the algebra (\ref{algebra}), and thus
$SU(1,1)$ is also a symmetry of fuzzy AdS$_2$.\\
In the commutative case $\theta \to 0$, the $X^i$ are
commuting coordinates and one can parametrize AdS$_2$ by

\begin{equation}
r = X^{-1} + X^1\,, \qquad
t = \frac lr X^0\,. \label{parametr}
\end{equation}

This leads to the induced metric (\ref{metricr}).

In the noncommutative case, (\ref{parametr}) suggests the definition

\begin{equation}
r = X^{-1} + X^1\,, \qquad t = \frac l2(r^{-1}X^0 + X^0r^{-1})\,,
\end{equation}

where we have introduced symmetrized products for $r^{-1}$ and $X^0$ so
that $t$ is
a Hermitian operator \cite{Ho:2001br}. From (\ref{algebra}) it follows
that the commutation relation for $t$ and $r$ is given by

\begin{equation}
[t, r] = i\theta\,. \label{commrel}
\end{equation}

Further evidence for (\ref{commrel}) was given in \cite{Ho:2000fy} by
considering closed strings in AdS$_2$. Besides, we note that (\ref{commrel})
is preserved by diffeomorphisms generated by the three Killing vectors of
the metric (\ref{metricr}).\\
Finally we observe that, writing the AdS$_2$ metric in a
conformally flat form by introducing $x = l^2/r$,

\begin{equation}
ds^2 = \frac{l^2}{x^2}(-dt^2 + dx^2)\,,
\end{equation}

the new coordinates obey the commutation relation

\begin{equation}
[x, t] = i\theta l^{-2} x^2\,,
\end{equation}

which is that of a quantum plane structure.

\subsection{Deformed solutions}

Besides the hermitian metric defined in equation
(\ref{metric}), there is another fundamental bitensor which can be
constructed in a natural way and corresponds to a nonantisymmetric volume
2-form:

\begin{equation}\label{volform}
E_{\mu\nu}=\epsilon_{ab}e^a_{\mu}\star e^b_{\nu}=H_{\mu\nu}+iM_{\mu\nu}\,,
\end{equation}

\noindent where

\begin{equation} \label{avolform}
H_{\mu\nu} = \frac 12 \epsilon_{ab}\{ e^a_\mu,e^b_{\nu}\}
\end{equation}

is real and antisymmetric and reduce to the usual volume form in the 
commutative case, whereas
 
\begin{equation} \label{svolform}
M_{\mu\nu} = -\frac i2 \epsilon_{ab}[ e^a_\mu,e^b_{\nu}]
\end{equation}

is real and symmetric and vanishes for $\theta^{\mu\nu}=0$.

We observe that, while in the commutative case the tensors $G_{\mu\nu}$
and $E_{\mu\nu}$ are invariant 
under the gauge transformations corresponding to the boost and the 
$U(1)$ gauge symmetry, in the noncommutative case this invariance property
is no longer valid. Using Eq.(\ref{boost}), it is easy 
to show that under an infinitesimal boost  these fields transform as

\begin{equation}
\delta_{\xi\tau_2}G_{\mu\nu} = -\frac 12 [E_{\mu\nu},\xi]\,, \qquad
\delta_{\xi\tau_2}E_{\mu\nu} = -\frac 12 [G_{\mu\nu},\xi]\,.
\end{equation}

The fields $g_{\mu\nu}$, $B_{\mu\nu}$, $H_{\mu\nu}$ and 
$M_{\mu\nu}$ transform according to

\begin{eqnarray}  \label{zero}
\delta_{\xi\tau_2} g_{\mu\nu} = -\frac i2 [M_{\mu\nu},\xi]\,, && \qquad
\delta_{\xi\tau_2} B_{\mu\nu} = \frac i2 [H_{\mu\nu},\xi]\,, \nonumber \\
\delta_{\xi\tau_2} H_{\mu\nu} = -\frac i2 [B_{\mu\nu},\xi]\,, && \qquad
\delta_{\xi\tau_2} M_{\mu\nu} = \frac i2 [g_{\mu\nu},\xi]\,.
\end{eqnarray}

\noindent For an infinitesimal $U(1)$ gauge transformation, we obtain
from Eq.(\ref{U(1)symmetry}) the transformation rules

\begin{equation}
\delta_{\chi\tau_3}G_{\mu\nu} = \frac i2 [G_{\mu\nu},\chi]\,, \qquad
\delta_{\chi\tau_3}E_{\mu\nu} = \frac i2 [E_{\mu\nu},\chi]\,,
\end{equation}

or equivalently

\begin{eqnarray}
\delta_{\chi\tau_3} g_{\mu\nu} = \frac i2 [g_{\mu\nu},\chi]\,, && \qquad
\delta_{\chi\tau_3} B_{\mu\nu} = \frac i2 [B_{\mu\nu},\chi]\,, \nonumber \\
\delta_{\chi\tau_3} H_{\mu\nu} = \frac i2 [H_{\mu\nu},\chi]\,, && \qquad
\delta_{\chi\tau_3} M_{\mu\nu} = \frac i2 [M_{\mu\nu},\chi]\,.
\end{eqnarray}

\noindent We see that in the noncommutative case all the fields
become charged under the $U(1)$ gauge symmetry.

These results can be summarized and generalized to finite gauge transformations
assembling $G_{\mu\nu}$ and $E_{\mu\nu}$ together in a doublet,

\begin{equation}
S_{\mu\nu}:= \left( \begin{array}{c} G_{\mu\nu} \\ E_{\mu\nu} 
\end{array}\right)=\left( \begin{array}{c} g_{\mu\nu} \\ H_{\mu\nu} 
\end{array}\right)+i\left( \begin{array}{c} B_{\mu\nu} \\ M_{\mu\nu} 
\end{array}\right)\,,
\end{equation}

which transforms under a finite boost or $U(1)$ gauge transformation as

\begin{equation}
S^T_{\mu\nu} \longmapsto \left[ U^{-1}\star S_{\mu\nu} \right]^T \star U\,.
\end{equation}

In order to deform the classical AdS$_2$ solution (\ref{metricr}),
we can iterate Eq.~(\ref{zero}) twice, obtaining, up to second
order in $\theta = \theta^{01}$

\eqn
g_{tt} &=& -\frac{r^2}{l^2} + \theta^2\frac{r}{2 l^2}
           [\partial_r\partial_t\xi\partial_t\xi - \partial^2_t\xi\partial_r
           \xi + r^{-1}\partial_t\xi\partial_t\xi]
           + {\cal O}(\theta^3)\,, \nonumber \\
g_{rr} &=& \frac{l^2}{r^2} + \theta^2\frac{l^2}{2r^3}[\partial_r\partial_t
           \xi\partial_t\xi - \partial^2_t\xi\partial_r\xi - 3r^{-1}
           \partial_t\xi\partial_t\xi]
           + {\cal O}(\theta^3)\,, \label{gMdef} \\
M_{tt} &=& -\theta\frac{r}{l^2}\partial_t\xi
           + {\cal O}(\theta^3)\,, \nonumber \\
M_{rr} &=& -\theta\frac{l^2}{r^3}\partial_t\xi
           + {\cal O}(\theta^3)\,, \nonumber
\feqn

where $\xi(t,r)$ is an arbitrary function, which should of course
be restricted appropriately if one wants the metric to approach
AdS$_2$ asymtotically, i.~e.~, for $r \to \infty$.
Note that the antisymmetric volume form $H_{\mu \nu}$, as well as
$B_{\mu \nu}$, continue to be zero.

We further observe that corrections to the metric become very large when
$r \rightarrow 0$,
so that perturbative calculations in the
noncommutativity parameter $\theta$ are not reliable near the horizon.

\section{Conclusions}

In this paper, we presented a model of noncommutative gravity in
two dimensions based on a deformation of an $SU(1,1)$ topological
gauge theory. The substitution of ordinary products by the Moyal
product makes it necessary to enlarge the gauge group to $U(1,1)$.
This amounts to introducing an additional abelian gauge field as
well as an additional scalar. Those fields corresponding to the
trace part are coupled to the gravitational fields in the
noncommutative case, and decouple for $\theta^{\mu\nu} = 0$.
We also showed that the deformed action admits an equivalent
formulation in terms of a matrix model,
whose coupling constant contains the noncommutativity parameter.

The metric that we defined contains a symmetric part, which reduces to the
ordinary metric once the noncommutativity parameter is set to
zero, and an antisymmetric part that vanishes for $\theta^{\mu\nu} = 0$.

Furthermore, some solutions of the noncommutative model, like
fuzzy AdS$_2$, were obtained, and symmetries of the deformed action
were studied. In particular, it was found that the action is
invariant under a class of transformations that reduce to ordinary
diffeomorphisms in the commutative case.

We saw that the Seiberg-Witten formula maps the topological gauge theory
on noncommutative spaces to the standard commutative one. This
behaviour is known from three-dimensional Chern-Simons
theory \cite{Grandi:2000av}, and might be related to the topological
nature of the model. One has to keep in mind, however, that the
Seiberg-Witten map is of perturbative nature in $\theta$, and
that the deformed action can admit solitonic solutions that become singular
when the noncommutativity parameter tends to zero. Therefore, we do not
expect the deformed model to be entirely equivalent to the undeformed one.

Further development of our work would be to address issues like
the quantization of the theory, or the classification of solutions of
the matrix model (\ref{matrixmodel}).
We hope to report on this in the near future.

\section*{Acknowledgements}
\small

This work was partially supported by INFN, MURST and
by the European Commission RTN program
HPRN-CT-2000-00131, in which the S.~C.~, D.~K.~, L.~M.~and D.~Z.~are
associated to the University of Torino.
\normalsize

\newpage

\begin{appendix}

\section{Conventions}

An element $M$ of the Lie algebra $u(1,1)$ satisfies

\begin{equation}
M^a_{\,\,\,b} = - \eta_{bc}\bar{M}^c_{\,\,\,d}\eta^{da}\,, \label{u11}
\end{equation}

where a bar denotes complex conjugation, and
$(\eta_{ab}) = {\mathrm{diag}}(-1,1)$. We choose as $u(1,1)$ generators

\begin{eqnarray}
\tau_0 &=& \frac 12\left(\begin{array}{cc} i & 0 \\ 0 & -i \end{array}
           \right)\,, \qquad
\tau_1 = \frac 12\left(\begin{array}{cc} 0 & 1 \\ 1 & 0 \end{array}\right)\,,
         \nonumber \\ && \nonumber \\
\tau_2 &=& \frac 12\left(\begin{array}{cc} 0 & -i \\ i & 0 \end{array}
           \right)\,, \qquad
\tau_3 = \frac 12\left(\begin{array}{cc} i & 0 \\ 0 & i \end{array}\right)\,.
         \label{generators}
\end{eqnarray}

They are normalized according to

\begin{equation}
{\mathrm{Tr}}(\tau_A\tau_B) = \frac 12 \eta_{AB}\,,
\end{equation}

where $(\eta_{AB}) = {\mathrm{diag}}(-1,1,1,-1)$ is the inner product
on the Lie algebra. The generators (\ref{generators}) satisfy the relation
(\ref{u11}). Further, if $i,j,k$ assume the values $0,1$ and $2$, then
the following relations hold:

\begin{eqnarray}
\left[ \tau_i ,\tau_j \right] &=& -\epsilon_{ijk}\tau^k \\
\left[ \tau_i ,\tau_3 \right] &=& 0\\
\tau_i \tau_j &=& -\frac{1}{2}\epsilon_{ijk}\tau^k - \frac{i}{2}
\eta_{ij} \tau_3 \\
{\mathrm{Tr}}(\tau_i\tau_j\tau_k) &=& -\frac 14 \epsilon_{ijk}\\
{\mathrm{Tr}}(\tau_i\tau_j\tau_3) &=& \frac i4 \eta_{ij}
\end{eqnarray}

where $(\eta_{ij}) = {\mathrm{diag}}(-1,1,1)$ and $\epsilon_{012} = 1$.\\
We furthermore defined $\epsilon_{01} = 1$, and
antisymmetrize with unit weight, e.~g.~

\begin{equation}
e^a_{\left[\mu\right.} \star e^b_{\left.\nu\right]} \equiv
\frac 12 \left(e^a_{\mu} \star e^b_{\nu} - e^a_{\nu} \star e^b_{\mu}\right)\,.
\end{equation}

\section{Gauge transformations of the gravitational fields}
\label{transformations}

In this appendix we write the explicit form of the different gauge 
transformation for the gravitational fields defined in Eq.(\ref{decomp}).
If we divide gauge transformations defined in Eq.~(\ref{gaugetransformations})
in {\em translations} 
($\lambda=\alpha^a \tau_a$), {\em boost} ($\lambda=\xi \tau_2$) and {\em 
$U(1)$ gauge symmetry} ($\lambda=\chi \tau_3$), 
their action on the fields turns out to be the following:

\begin{itemize}

\item Translations

\begin{eqnarray} \label{translations} 
\delta_{\alpha}e^a &=& l d\alpha^a + \frac l2 \epsilon^a{}_b\{\omega,
\alpha^b\} + \frac{il}{2}[b,\alpha^a]\,, \nonumber \\
\delta_{\alpha}\phi^a &=& \frac 12 \epsilon^a{}_b\{\phi,\alpha^b\} +
\frac i2 [\rho,\alpha^a]\,, \\
\delta_{\alpha}\omega &=& -\frac{1}{2l}\epsilon_{ab}
\{ e^a,\alpha^b\}\,, \qquad
\delta_{\alpha}\phi = -\frac 12 \epsilon_{ab}\{\phi^a,\alpha^b\}\,,
\nonumber \\
\delta_{\alpha}b &=& -\frac{i}{2l} \eta_{ab}[e^a, \alpha^b]\,,
\qquad
\delta_{\alpha}\rho = -\frac{i}{2l} \eta_{ab}[\phi^a, \alpha^b]\,.
\nonumber
\end{eqnarray}

\item Boost

\begin{eqnarray} \label{boost}
\delta_{\xi}e^a = -\frac 12 \epsilon^a{}_b\{e^b,\xi\}\,, && \qquad
\delta_{\xi}\phi^a = -\frac 12 \epsilon^a{}_b\{\phi^b,\xi\}\,,
\nonumber \\
\delta_{\xi}\omega = d\xi + \frac i2 [b,\xi]\,, && \qquad
\delta_{\xi}\phi = \frac i2 [\rho,\xi]\,, \\
\delta_{\xi}b = -\frac i2 [\omega, \xi]\,, && \qquad
\delta_{\xi}\rho = -\frac i2 [\phi, \xi]\,. \nonumber 
\end{eqnarray}

\item U(1) gauge symmetry

\begin{eqnarray} \label{U(1)symmetry} 
\delta_{\chi}e^a = \frac i2 [e^a,\chi]\,, && \qquad
\delta_{\chi}\phi^a = \frac i2 [\phi^a,\chi]\,, \nonumber \\
\delta_{\chi}\omega = \frac i2 [\omega,\chi]\,, && \qquad
\delta_{\chi}\phi = \frac i2 [\phi,\chi]\,, \\
\delta_{\chi}b = d\chi +\frac i2 [b, \chi]\,, && \qquad
\delta_{\chi}\rho = \frac i2 [\rho, \chi]\,. \nonumber
\end{eqnarray}

\end{itemize}

\section{Reality of the vector field $v$}

\label{vreal}

We still have to prove that the vector field $v$
generating the transformations (\ref{defdiff})
can always be chosen to be real.
Indeed assuming the zweibein $e^a_{\mu}$ to be invertible
as an ordinary matrix, we look for a 
solution $v$
of $l\alpha^a=i^\star_{v}e^a=v^\mu\star_S 
e_\mu^a$, with
$\star_S:= \cos(\frac 12 \bide)$,
written as a series in $\theta^{\mu\nu}$,

\eqn
v=\sum_{n=0}^{\infty} \stackrel{(n)}{v}\ ,
\feqn 

where $\stackrel{(n)}{v}$ is of order $n$ in $\theta^{\mu\nu}$.

Now we write

\eqn \label{condition}
l\alpha^a &=& v^\mu\star_S e^a_\mu \cr
&=& \sum_{n,m=0}^\infty \frac{(-)^m}{(2m)!}
\stackrel{(n)}{v}\!\!{}^\mu \big(\frac 12 \bide\big)^{2m} e^a_\mu \cr
&=& \sum_{n,m=0}^\infty \frac{(-)^m}{(2m)!}\Big[
\stackrel{(2n)}{v}\!\!{}^\mu \big(\frac 12 \bide\big)^{2m} e^a_\mu
+\stackrel{(2n+1)}{v}\!\!{}^\mu \big(\frac 12 \bide\big)^{2m} e^a_\mu\Big]
\cr
&=& \sum_{k=0}^\infty \sum_{m=0}^k \frac{(-)^m}{(2m)!}\Big[
\stackrel{(2k -2m)}{v}\!\!{}^\mu \big(\frac 12 \bide\big)^{2m} e^a_\mu
+\stackrel{(2k-2m+1)}{a}\!\!{}^\mu \big(\frac 12 \bide\big)^{2m} 
e^a_\mu\Big]\,.
\feqn

We find that the zero 
order condition admits a unique solution

\eqn
\stackrel{(0)}{v}\!\!{}_a^\mu e^a_\mu=\alpha^a\,,
\feqn

whereas the first order condition

\eqn
\stackrel{(1)}{v}\!\!{}^\mu e^a_\mu=0
\feqn

implies that $\stackrel{(1)}{v}\!\!{}^\mu\equiv 0$. By induction, knowing
$\stackrel{(n)}{v}\!\!{}^\mu$ for every  $n\leq 2k$, the  $\stackrel{(2k+2)}{v}\!\!{}^\mu$ and
$\stackrel{(2k+3)}{v}\!\!{}^\mu$ are
uniquely determined in terms respectively of the  
$\stackrel{(2n)}{v}\!\!{}^\mu$ and
$\stackrel{(2n+1)}{v}\!\!{}^\mu$ of 
lower order. So, if $\stackrel{(1)}{v}\!\!{}^\mu\equiv 0$, then
$\stackrel{(2n+1)}{v}\!\!{}^\mu\equiv 0$ for every  $n$ and knowing
$\stackrel{(0)}{v}\!\!{}^\mu$ one can construct iteratively the complete
solution of the condition (\ref{condition}).

\section{Transformation properties of the fields}

\label{transfprop}

Under the action of the transformations (\ref{ncdiff}), the
fields (\ref{decomp}) transform as

\begin{eqnarray}
\Delta_v^\star e^a_\mu &=& {\cal L}^\star_v e^a_\mu + \frac 14 
\epsilon^a{}_b
\big( [\omega_\nu,[e^b_\mu,v^\nu]]-[e^b_\nu,[b_\mu,v^\nu]]\big) \cr
&& + \frac i4 \big( \{ e^a_\nu,[b_\mu,v^\nu]\} + \{ b_\nu , [e^a_\mu ,v^\nu]\}
\big)\ ,\cr
&&\cr
\Delta_v^\star \omega_\mu &=& {\cal L}^\star_v \omega_\mu -
\frac{1}{4l^2} \epsilon_{ab}
[e^a_\nu,[e^b_\mu,v^\nu]] \cr
&& + \frac i4 \big( \{\omega_\nu,[b_\mu,v^\nu]\} + \{b_\nu , [\omega_\mu,
v^\nu]\} \big)\ ,\cr
&&\cr
\Delta_v^\star b_\mu &=& {\cal L}^\star_v b_\mu -\frac{i}{4l^2}\big(\eta_{ab}
\{ e^a_\nu,[e^b_\mu,v^\nu]\} + \{\omega_\nu,[\omega_\mu,v^\nu]\} \big) \cr
&& + \frac i4 \{ b_\nu,[b_\mu,v^\nu]\}\ ,\cr
&&\cr
\Delta_v^\star \phi^a &=& {\cal L}^\star_v \phi^a + \frac 14 \epsilon^a{}_b
\big( [\omega_\nu,[\phi^b,v^\nu]] - \frac 1l [e^b_\nu,[\phi,v^\nu]]\big) \cr
&& + \frac i4\big(\frac 1l \{ e^a_\nu,[\rho,v^\nu]\} + \{ b_\nu ,
[\phi^a ,v^\nu]\}
\big)\ ,\cr
&&\cr
\Delta_v^\star \phi &=& {\cal L}^\star_v \omega_\mu -
\frac{1}{4l} \epsilon_{ab}
[e^a_\nu,[\phi^b,v^\nu]] \cr
&& + \frac i4 \big(\{\omega_\nu,[\rho,v^\nu]\} + \{b_\nu, [\phi,v^\nu]\}
\big)\ ,\cr
&&\cr
\Delta_v^\star \rho &=& {\cal L}^\star_v \rho - \frac i4 \big(  
\frac 1l \eta_{ab}
\{ e^a_\nu,[\phi^b,v^\nu]\} + \{\omega_\nu,[\phi,v^\nu]\} \big) \cr
&& + \frac i4 \{ b_\nu,[\rho,v^\nu]\}\ .
\end{eqnarray}

\end{appendix}

\end{document}